\documentclass{aa}
\usepackage{graphics}

\begin{document}
\def\etal{{\it et al.\ }}
\def\vx{\hbox{\bf x}}
\newcommand{\asb}[3]{#1, Ann. sco. sci. Bruxelles, #2\rm, #3}
\newcommand{\nat}[3]{#1,  Nature, #2\rm, #3}
\newcommand{\np}[3]{#1,  Nucl. Phys.,  #2\rm, #3}
\newcommand{\apj}[3]{#1,  ApJ,  #2\rm, #3}
\newcommand{\aj}[3]{#1,  AJ,  #2\rm, #3}
\newcommand{\apjl}[3]{#1, ApJ,  #2\rm, L#3}
\newcommand{\apjsup}[3]{#1,  ApJS, #2\rm, #3}
\newcommand{\aeta}[3]{#1, A\&A,  #2\rm, #3}
\newcommand{\annrev}[3]{#1, ARA\&A,  #2\rm, #3}
\newcommand{\mnras}[3]{#1,  MNRAS, #2\rm, #3}
\newcommand{\jrasc}[3]{#1,  JRASC,  #2\rm, #3}
\newcommand{\pasj}[3]{#1,  PASJ,  #2\rm, #3}
\newcommand{\sa}[3]{#1,  Sv A,  #2\rm, #3}
\newcommand{\physlettb}[3]{#1, Phys. Lett. B, #2\rm, #3}
\newcommand{\physrevlett}[3]{#1, Phys. Rev. Lett., #2\rm, #3}

%\thesaurus{02        % A&A Section 2: Cosmology  (03.12.1  % Cosmology      )}

\title{Warm photo-ionized IGM: a Clue for galaxy and cluster formation history?
}
\author{S.Prunet$^{1,2}$  
\and A. Blanchard$^{3,4}$}
\offprints{S.Prunet}
\institute{ $^1$IAS, Universit\'e de Paris XI, B\^at 121, F--91405 Orsay Cedex\\
$^2$CITA, University of Toronto, 60 St George Street, Toronto ON M5R3H8, Canada\\
$^3$Observatoire Astronomique de Strasbourg, 11 rue de l'Universit\'e,
F--67000 Strasbourg
\\
$^4$Observatoire Midi-Pyr\'en\'ees, LAT, 31 Av. Ed.Belin, 31~400 Toulouse}
\maketitle
\begin{abstract}
In this paper we investigate  the overcooling problem and propose some 
possible solutions. We show that the overcooling problem is
generic to the hierarchical picture of structure formation, as long 
as the cooling is actually possible
 in small halos at high redshift. Solutions to this problem are likely to be 
associated with some feedback mechanism, and probably imply the existence of a 
warm IGM containing most of the cosmological baryons.
We concentrate on some possible solutions, mainly photoionization and 
bulk-heating of the 
IGM. We show photoionization can act as a significant 
feedback mechanism but is not strong enough to solve the entire overcooling 
problem.  We therefore assume that the IGM is maintained hot by some energy injection 
provided by supernova and galaxy formation is then limited by this feedback mechanism. 
Such a self regulated scheme allows us to compute the thermal 
history of the IGM. In the absence of any photo-ionization, we find the  
existence of bifurcations in 
the thermal history. However, these bifurcations are suppressed 
when significant photo-ionization occur in addition to the energy injection. 
The temperature of the IGM is then maintained at $T \sim $ few $10^{5}$K.
We find that for realistic fraction of the energy produced by supernovae 
being 
re-injected into the IGM, this scenario can consistently reproduce the present 
amount of stars. A $\Gamma-$CDM model with $\Gamma\sim 0.25$ can reproduce 
properly the observed HI gas at high redshift. A simple prescription for star 
formation also allows to reproduce the star formation rate at high redshift 
as inferred from recent data, while the entropy of the IGM at $z \sim 2-5$ is 
of the order of what seems necessary to explain the observed properties of X-ray clusters. 
We conclude that the warm self-regulated IGM picture provides
an interesting alternative to standard semi-analytical approach, which may 
elucidate the behavior of baryons in structures from small galaxies to 
clusters.

\end{abstract}

\section{Introduction}

The global picture of the formation of structures on all scales
is one of the widest questions in modern cosmology. Even the validity
of the gravitational instability paradigm has not yet received 
complete confirmation and the door remains open to alternative 
approaches, like the topological defect models 
(see for instance Durrer \etal 1999). However,
the gravitational instability picture is the most compelling one,
primarily since it can be investigated to an unprecedented level
of detail. The small perturbations
necessary to seed structure formation are believed to originate in the very early
universe. The obvious non-linear character of structures like 
clusters and galaxies has lead to the intensive 
use of numerical simulations
which are now achieving remarkable performances. In this context , the r\^ole of baryons has also 
been included in recent years by incorporating the hydrodynamics of gas in the simulations. 
However, analytic modeling is also essential to our understanding of the physics, 
even if some aspects require
the validation by numerical simulation. The clearest 
example of this is the mass function as inferred by 
Press and Schechter: it is only since numerical simulations
have confirmed its validity, at least as an efficient fit to the
actual mass function arising in numerical simulations, that 
it has been widely used for cosmological applications. 
Cosmological constraints that can be inferred from 
properties of clusters is certainly the field which has the most 
benefited of this approach, although it is not yet entirely
clear whether we understand the physics of the gas in clusters. For instance,
the relation between luminosity and temperature is not what one
would expect in a simple scaling model as proposed by Kaiser
(1986): the origin of the problem probably lies in the fact that the formation
process of the X-ray core is not well understood (Blanchard \& Silk 
1990) possibly requiring non-gravitational processes (Evrard 
\& Henry 1991; Kaiser  1991).
However, the impressive agreement of the Press and Schechter formula with the
mass function on every scale, for which some theoretical reasons
exist, has lead to handle the question of structure formation on 
galactic and sub-galactic scales up to the question 
of the first structures that formed
(Tegmark \etal 1997). On galactic scales it is clear that 
dissipative processes play a key r\^ole, as they are necessary
to allow star formation. It has been proposed that the cooling criterion is
an essential requirement for galaxy formation: in order to form a galaxy
it seems actually unavailable that the baryons can dissipate
their thermal energy and reach a density high-enough that star formation
can actually start, although the details of this process are still
far from being understood: even star formation in our own Galaxy or in 
neighboring galaxies is not well understood. The fact that first stars 
should form 
in the absence of any dust makes the situation more problematic. The inclusion 
of the baryonic component in PS halos was pioneered by White and Rees (1978). 
They tried to infer the shape  of  the  luminosity function of galaxies, and noticed that 
it was expected  to be steeper than the mass function at the faint end, leading to a  
luminosity function  far too steep compared to observations. This was also noticed 
by Peacock and Heavens
(1990)  in the context of the peak formalism approach to the mass function. The reality of 
this problem has become obvious, in more recent analysis, in which several 
authors have tried to infer the
shape of the luminosity function in the context of the CDM theory (White \& Frenk 1991; 
Lacey \& Silk 1991).
In their approach, White and Frenk (1991) developed a model  for the formation and 
evolution of galaxies
in which  a lot of energy was injected  into the gas which has felt in the dark matter 
potential wells,
something which was achieved at the price of a metal production level  which 
appeared unrealistically large. 
Since, similar models have been  build in order to achieve
a better agreement with  observations (Kauffmann \etal 1993; Cole \etal 1994). 
In the context of the Cold Dark Matter picture in an Einstein de Sitter universe Blanchard \etal
(1990; 1992a), BVM hereafter, investigated the cosmological history of the cooling of the gas. They concluded
that the simple cooling argument leads to the conclusion that  most of baryonic gas was likely  
 to have been cooled at some time of its history,
leading to the so called {\em overcooling} problem: the amount of gas which would be expected today 
to be in cold and probably dense state would represent more than 80\% 
of the cosmological baryonic content of the universe, one order of magnitude  
larger
 than the baryonic content of galaxies. This problem was also noticed independently by Cole  (1991). 
The overcooling problem is a great challenge to numerical simulations as spatial 
resolution of $10^3$ is 
necessary and a mass resolution of $10^6$ 
(first significant cooling occurs in objects which have a virial temperature of $10^4$K at $Z \sim 10$ corresponding to mass of $10^8$ M$_\odot$ with typical size of 1 kpc). However, the high resolution 
numerical simulations of 
Navarro and Steinmetz (1997) were able to see this overcooling problem, with 
an 
amount of cooled gas which agrees quite well with a simple argument based 
on Press and Schechter estimate.
This gives enough confidence in the analytical approaches for further 
investigations of the 
problem, and  its possible solutions. Indeed, any reliable scenario for 
galaxy formation
should solve this problem in a self-consistent way. BVM
proposed that the IGM is heated by the feedback 
of first structures  to temperature high enough (typically a few $10^5$K)
 to prevent further collapse in most potential wells. In this picture most 
of the IGM stays 
homogeneous. 
This warm  IGM picture was shown to potentially offer an elegant solution 
to the origin of the shape
of the luminosity function, leading to an inverse hierarchical picture, 
in which large galaxies
form at high redshift while at least some 
small galaxies form at low redshift. One of the 
objectives of the present work is to investigate this idea in a more 
quantitative way. Furthermore, several 
new observations provide information relevant to galaxy formation: 
the CFRS redshift survey 
has illustrated the apparent increase of the  star formation rate up to  
redshift of the order of one,
 while the  HDF seems to reveal a decrease of the star formation 
rate at higher $z$. Dust obscuration may however occur at large redshift, masking a significant fraction of star formation. Furthermore, the 
first field galaxies at redshift as high as 4 have been discovered 
(Steidel 1996) and  the neutral 
gas content in dense objects  has been  evaluated up to redshift 4, 
this gas potentially being the precursors
of disks. It is therefore interesting to see if such observations 
can be accommodated by this model.
In the first section, we examine the overcooling problem  in various 
models and its sensitivity to the various input parameters.
We then examine the possible r\^ole of photo-ionization and show that 
photo-ionization implies 
a significantly smaller fraction of cooled gas, but is still unable 
to solve the overcooling 
problem. 
We then investigate in some detail the original scenario of  BVM in 
which the IGM 
is heated by the feedback of galaxy formation, resulting in a self 
regulated evolution of the 
temperature of the IGM. This model will be called the warm IGM picture 
hereafter. The basic conclusion 
of our study is that the warm IGM picture  possesses few parameters but 
can easily describe several global 
properties of the cosmological baryonic history that can be inferred from 
the observations. It may therefore represents an interesting alternative to 
more 
traditional semi-analytical scenarios.

\section{The  Overcooling problem}

The physics of the formation of  galaxies is still rather unknown. However, it is almost trivial to 
say that the  physics of baryons should play an essential role. For instance, the origin of the 
characteristic size, mass and luminosity of a bright galaxy has been first 
looked for in the physics of the
linear evolution of fluctuations before recombination or after the recombination (see for instance 
Gamov 1948). However, the Jeans masses after and before 
recombination  were recognized as being very different from the typical galaxy mass, 
while the damping leads to masses too large 
to be identified with galaxies  (Silk 1967). One key criterion which has been proposed
 to define the scale associated with typical galaxies (say an $L_{*}$ galaxy) 
is based on the cooling criteria:
  during  the gravitational collapse of an object the kinetic  energy of the gas is turned into thermal 
energy, which is then lost by some mechanism in order for  the gas to 
contract further 
up to densities where star formation has any reasonable chance to start (of 
course, this is a
minimal requirement, as the mechanism of star formation may 
require many more 
conditions!). It has 
been suggested that 
$L_{*}$ galaxies 
are the largest objects able to cool in a time scale shorter than the 
typical age of the universe
(Binney 1997; Rees \& Ostriker 1977; Silk 1977). 
These kinds of considerations were already investigated by Hoyle as early as 
1953 (Hoyle, 1953). 

It is in this context that the {\em overcooling} problem has been outlined, which is likely to occur
in any hierarchical model of galaxy formation. The nature of the overcooling has been explicitly 
stated  for the first time by Blanchard \etal (1990; 1992a) and Cole (1991).  In the simplest picture,
one assumes that 
the gas which cools is turned into stars,  so the total amount of cooled gas should represent the 
present day amount of stars, this is quite reasonable as the amount of cold gas (HI) at low redshift is modest.  In the context of hierarchical models, 
it is expected that at some redshift a large fraction of the matter 
of the universe settles into small potential wells with circular velocity between 20 and 200 km/s, 
the virial temperature of the gas being then at temperature higher than $10^4$K and at rather high
 density. 
Under these conditions  the radiative  cooling is very efficient and therefore 
the cooling time is extremely short. One then naturally expects a large fraction of the cosmological gas to be cooled
by now. This is in contradiction with a standard result in Cosmology:
standard nucleosynthesis value for the baryonic  density parameter of the universe
is of the order of $0.08 h_{50}^{-2}$ (Burles \etal 1999), while the present density in stars 
There are several considerations which are important to take into account  
 while dealing with the overcooling problem: using the cooling criterion 
as a criterion
for star formation reflects our poor knowledge of the realistic physical conditions 
required for star formation. 
However, the cooling criterion seems clearly to be a minimal requirement, 
and it is likely to be valid  
anyway in order to evaluate the total amount of 
cold gas, counting  stars in this category. 
Because the present amount 
of cold dense gas actually seen in the universe is rather small, it is 
reasonable to imagine that 
cold gas has been actually turned into stars. Some alternatives are possible 
though: there might be 
more gas than what we see (Pfenniger \& Combes 1994), or most of the cold gas may 
have been turn into stars
which are not seen. In both cases, this represents a way to generate 
candidates for baryonic 
dark matter. In such a case,  the  overcooling question  would be intimately linked
to the solution of the 
dark matter problem! This could at least point out to a possible origin of 
the machos in galactic halo.
Machos,  if they exist,  may provide evidence that a substantial fraction of 
the baryonic density of the universe
is dark, at least  if the situation in our Galaxy can be taken as 
representative of the general situation (see Alcock \etal 1998). 
One could also 
wonder if the apparent overcooling problem is not due to the fact that
 nucleosynthesis 
estimates are erroneous. However,  this can not  provide a satisfying answer.
 The reason comes from the x-ray properties of clusters : in cold dark matter picture the baryonic content
 is 
expected to be the sum of the stars in galaxies and of the x-ray gas, the 
latter representing at least 
80\%$h^{-3/2}$
of the total baryonic content, meaning that the fraction of gas which is 
in the ``cool'' phase 
(stars in this case)  is a small fraction of the total baryonic content. More 
precisely,
the typical ratio of the baryonic mass seen in clusters to the mass in 
stars is of the order of 
5-6 $h^{-3/2}$ (White \etal 1993), so assuming that this ratio is universal 
leads to an estimate of the baryonic content of 
the universe, mainly in the form of intergalactic gas :
\begin{equation}
\Omega_{baryon\leftarrow cluster} \approx  (1 + 5-6 h^{-3/2}) \Omega_{\star} 
\sim 0.03-0.06 h^{-3/2}
\end{equation}
interestingly close  to primordial nucleosynthesis value 
$\sim 0.02h^{-3/2}$ for a low Deuterium abundance. Incidentally, this has  
interesting consequences: this value conflicts with 
nucleosynthesis values corresponding to high Deuterium abundance (Songaila \etal 1997)
and does not allow a significant contribution of machos to the cosmic baryon budget.
The agreement between the inferred value of $\Omega_b$ and nucleosynthesis reinforce 
the so-called baryons crisis (White \etal 1993), which is a 
problem specific to 
 $\Omega = 1$ models. 
Finally, the ratio of mass in visible stars to gas mass 
in clusters is consistent with the ratio 
$F_*$, the ratio  between $\Omega_*$ and $\Omega_{bbn}$ in the universe:
\begin{equation}
F_*  \sim 0.05-0.20
\end{equation}
This essentially means that only a small fraction of the primordial baryons 
were  turned into visible stars and that there is little room for an other 
significant dark baryonic component.

\subsection{Framework}

There has been a lot of progress in the investigation of cosmic scenarios for structure formation since it
has been realized, thanks to numerical simulations (Efstathiou \etal 1988; Cen, Gnedin \& Ostriker 1993;
Navarro, Frenk \& White 1996), that the Press and Schechter 
(1974) formula, (PS hereafter), for the mass function is a good working approximation. Our understanding 
of the behavior of the baryonic content has also been greatly improved by numerical simulations in which 
the fate of the cosmological gas is followed. Despite possible complex behavior, it seems well 
established that the gas is heated during the collapse either by shocks or by adiabatic compression, 
until it reaches the virial temperature at which  it can settle in nearly 
hydrostatic equilibrium:
\begin{equation}
T_v \sim 5\times 10^5 M_{12}^{2/3}(1+z)\, K
\end{equation}
 However, hydrostatic equilibrium is possible only as long as the gas 
does not loose a substantial fraction of its internal energy, i.e. during a 
period  of time 
smaller than  the typical cooling time. The PS mass function changes over a 
time scale comparable to the  
 age of the universe, so if the typical cooling time is longer than the age of the universe, the gas 
will be merged in an other structure before it can cool. On the contrary, if 
the gas can cool 
on a time scale much shorter than the age of the universe, it is unlikely that 
gravitational shaking will re-heat
the gas before it has cooled down. As soon as the gas can cool 
it will contract further.
The  gas will be re-heated during the contraction phase 
by adiabatic compression, 
however as the 
cooling time becomes  shorter with increasing density there is a runaway 
catastrophe, which can apparently 
be stopped only when the gas has dissipated its thermal energy, probably 
being then  
rotationally supported or having been turned into stars. 
  
\begin{figure}
\resizebox{\hsize}{!}{\includegraphics{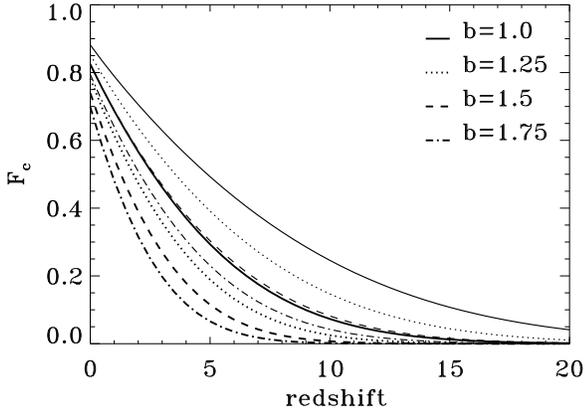}}
\caption{\label{fig:Fi1} In this figure we have estimated the integrated 
fraction of gas $F_c$  (using Eq \ref{eq:Fi2}) in two $\Gamma-$CDM pictures 
with $\Gamma = 0.5$ (thin lines)
and $\Gamma =0.25$ (hick lines) with different values of the bias parameter with 
$\eta = 1$. As one can see the history of the cooling depends  
on the parameters of the model, but the present day total amount of baryons 
that have cooled is relatively independent of the model, and 
reaches a large value, close around  80\%. }
\end{figure}

\begin{figure}  %[htbp]
\resizebox{\hsize}{!}{\includegraphics{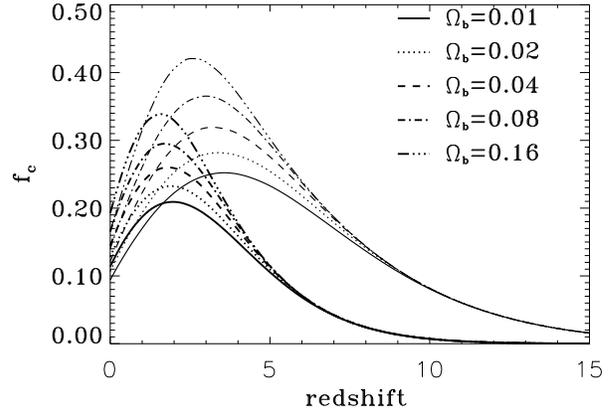}}
\caption{\label{fig:gzomb} On this figure we have computed the instantaneous 
fraction of gas $f_c$  in two $\Gamma-$CDM pictures with $\Gamma = 0.5$ and 
$\Gamma =0.25$ with different values of the baryonic content of the universe 
with $\eta = 1$ and $b = 1.5$.}
\end{figure}
% CDM b = 1.5   

%\subsection{sensitivity to parameters}
 The instantaneous fraction of gas able to cool  can be computed as a simple integral:
\begin{equation}
f_c(z) = \frac{1}{\rho}\int_{m_1}^{m_2} N(m)m w(m)dm
\end{equation}
where $w(m)$ represents the fraction of the baryonic material present in the 
structure $m$ which is able to cool or has already cooled at an earlier epoch. 
BVM showed that $w(m)$
 can be evaluated by assuming that all the baryonic content 
is in a homogeneous phase  inside the halo and estimating its cooling  time: if 
this cooling time $t_c(m)$ is shorter than some characteristic  time $t_1$, the gas 
will cool before undergoing any substantial re-heating.  Such a characteristic 
time scale will always be a fraction of the Hubble time at the considered 
redshift:
\begin{equation}
t_1  = \eta t_H
\end{equation}
with $\eta$ smaller than 1, and that can be taken to be independent of the mass. 
This may not be strictly valid: the typical time scale may depend on the 
power spectrum and on the  amplitude of the initial fluctuations associated. 
However, it is doubtful that the effect is important.  
Accordingly to BVM, we assume  
 that the cooling fraction is either 100\% or 0\% i.e.  $\omega(m)$ can take only two values (0 or 1) and  cooling is efficient only for masses in some  range
 $m_1 -m_2$.
The instantaneous cooling fraction can then be written as:
\begin{equation}
f_c(z) = \frac{1}{\rho}\int_{ m_1}^{m_2} m N(m) dm
\label{eq:fc2}
\end{equation}
The next issue is to estimate the total  fraction of cooled gas integrated 
over redshift. As long as the integrated fraction remain small one can writes:
\begin{equation}
F_c(z) \approx \int_{z}^{+\infty} \frac{f_c(z)}{\eta t_H} dt
\label{eq:Fi0}
\end{equation}
where $\eta t_H$ is a typical time scale for haloes to rearrange themselves.
 When the integrated fraction starts to become large this formula cannot be 
used any more: it becomes likely that most of the baryonic content of the 
halos in the cooling region has already cooled in a previous generation of 
halos. In principle, one has to follow the detailed history at all epochs of 
the precursors of these haloes, which could be  done for instance 
by using a merging 
history tree. However, BVM showed that the result of this computation can be evaluated more directly by taking into account the past history of the halos: large masses at low redshift are essentially 
built  from structures which were inside the cooling 
region at some higher redshift. Therefore, they should already contain only 
cooled gas. This leads to:
\begin{equation}
F_c(z) = \frac{1}{\rho}\int_{m_1}^{+\infty } N(m)mdm
\label{eq:Fi2}
\end{equation}
avoiding to have to handle the merging tree history. 
Numerical simulations indicate that this is a good working approximation
(Navarro \& Steinmetz 1997). In a 
realistic scenario however $F_c(z)$  has to remain small and the approximation 
\ref{eq:Fi0} can be used.

Let us now examine the typical amplitude one gets for the amount of cooling at various redshifts
 and the sensitivity to the various parameters. 
We have  evaluated the integrated fraction $F_c$ (Eq. \ref{eq:Fi2}) for  two CDM-like  
models with shape parameter \footnote{$\Gamma$, the shape parameter, is defined as
the apparent magnitude of $\Omega h$ in the CDM transfer function, see e.g. 
Peacock \& Dodds 1994} $\Gamma = 0.5$ and $\Gamma = 0.25$ (henceforth referred
to as $\Gamma$-CDM models)  with  various values of 
the normalization. 
This is presented in figure \ref{fig:Fi1}. Here and throughout the article
we adopted the cooling function with zero metallicity as given by Sutherland \& 
Dopita 1993. This figure illustrates the 
overcooling in a dramatic way: at redshift $z = 0$ more than 80\% of the 
primordial baryons should have been cooled. This is almost independent on the  
value of 
$\Gamma$ and on the the normalization $\sigma_8 = 1/b $, although the redshift 
history is sensitive to the specific model.

Because the amplitude of the overcooling problem is nearly a factor of ten and because the final value is rather insensitive to the detailed values of the parameters entering the calculation, 
it is not expected to be solved by adjusting one of the various parameters 
entering the problem. Examination of the instantaneous fraction (figures 
\ref{fig:gzomb} reinforces the idea that the problem is real:
there are epochs at which the {\em instantaneous amount} of gas able to cool
is in the range 20\% -- 50\% (the maximum value of $f_c$ gives a 
robust   
lower bound for the integrated value $F_c$). This  is already 
larger than what is implied by the observed amount of stars:  the  estimated present-day 
value is $F_* \sim 10\%$,
 Varying the baryonic content of the universe or the value of $\Gamma$ of the power spectrum 
or the normalization of
the models does not change significantly the maximum value of $f_c$. 
This illustrates  the significance of the overcooling problem.  
Another interesting consequence is that the 
amount of cold gas available at high redshift is high enough that most of 
the models can accommodate a main galaxy formation epoch as early as  $z\geq 5$: 
in fact in the context of the standard cold dark matter picture (with $b \sim 2$), 
all the observed stars could have formed at $z > 5$ ! 
The Canada-France Redshift Survey (CFRS) has provided evidence that a significant 
amount of star  formation 
has occurred since $z = 1$ (Lilly \etal 1995; 1996) and the Hubble Deep Field (HDF) may 
provide evidence  that the bulk of 
star formation occurs at low redshift, possibly as low as $z=1$ (Madau \etal 1996; 1997). 
A qualitative comparison of
 the instantaneous fraction $f_c(z)$ and of the  star formation rate as 
inferred from observations is instructive: from $f_c(z)$, one expects 
the bulk of star formation to occur at rather low redshift ($z_f \sim 1-4$). In other words, 
although the amount of gas available for star formation  is much too large 
compared to the  SFR, the shapes of the redshift distributions of both 
quantities are not very different (although a very late galaxy formation epoch, like $z \leq 1$, would not appear natural). This is consistent with the idea that 
galaxy formation has occurred in a self-regulated way by some mechanism, 
although this mechanism is not explicitly stated in most semi-analytic models. 
As argued by BVM, because such a mechanism should essentially 
limit the amount of star 
formation, it is reasonable to think that the same mechanism is 
responsible for the solution of the overcooling problem and might explain 
the shape of the luminosity function of galaxies. The scenario 
we investigate in the next section  is an example of such a scheme. \\

\section{Re-heated IGM and structure formation}

Although there is little doubt that re-heating is a fundamental problem
during galaxy formation, it is much harder to identify which 
astrophysical sources could be responsible for it, and to specify the detailed 
physics of the feedback mechanism. 

Following earlier consideration by Larson (1974), Dekel and Silk (1986)
 argued 
that star formation was self limited in small potentials by supernova
heating: in this model gas is heated and expelled from galaxies.
The same mechanism was advocated by Cole (1991). This effect is expected
to be significant for dwarf galaxies with masses $M \la 10^7 M_{\sun}$ 
(see Mac Low \& Ferrara, 1999). These authors also showed that if gas cannot be expelled
from larger galaxies, metals can still be expelled efficiently for masses up
to $M\simeq 10^8 M_{\sun})$.
Tegmark \etal (1993) have reexamined such a model
and concluded that the IGM could be re-ionized by supernova-driven winds, 
solving the Gunn-Peterson test, but probably producing a substantial 
amount of metals
at redshift $z= 5$ or larger.

A feedback mechanism has been incorporated into most recent 
models of galaxy formation, although it is generally implemented by assuming
that galaxy formation is inefficient within small halos, while the detailed physics
is not explicitly taken into account. This approach is the one 
used in several so-called semi-analytic models (eg Kauffmann \etal 1993;
Cole \etal 1994). In our opinion, this 
procedure might hide one of the most critical aspects: it is far from
 obvious that the overcooling problem has been solved then, even if the global 
modeling seems to match observations, because the physics of the feedback 
mechanism is not explicitly treated. There are basically two different kinds 
of feedback mechanisms. The first family is the one used by White and Frenk 
(1991), and is local in nature. The second advocated by BVM is global. 
The suppression of cooling by photo-ionization enters this last category.
Certainly one of the central issues of galaxy formation is to understand
whether the feedback mechanism which actually solves the overcooling problem is 
{\em local} or {\em global}. 

In the re-heated picture, the physical state of the IGM might be quite 
inhomogeneous and complex. In the scenario which 
we will consider we simplify the general picture by assuming that 
there are essentially two phases: one which is the condensate phase, 
corresponding to the gas which has been able to cool and can eventually  
be turned into stars,  and the IGM phase, which is supposed to be 
essentially homogeneous. At high redshift, as soon as the IGM is 
ionized, its cooling time is shorter 
than the Hubble time and it lies well inside  the 
cooling phase (unless its temperature is very high, but then this would 
produce unacceptably large Compton distortion in the CMB spectrum, given
the constraints provided by the COBE data, see Fixsen \etal 1997)
  and some heating process is necessary to maintain it in this hot phase. 

\subsection{Structure formation suppression mechanisms }

In order to solve the overcooling problem one needs to satisfy two 
constraints:
the integrated density of stars should not be larger than what is observed
in present-day visible stars and the amount of cooled gas at high redshift
should not be larger than what has been estimated from Damped Lyman alpha 
systems. Although there are ways to escape these constraints, in the following
we will investigate which scenarios naturally satisfy these constraints.\\

In order to reduce the amount of cooling that might happen during the 
cosmological history of baryons, there are two interesting mechanisms that one 
can think of. The first mechanism has been advocated by BVM: if the IGM is 
hot with a typical temperature $T_{{\rm IGM}}$ then structure formation will
be suppressed on scales for which the virial temperature is smaller than this:
$$
T_V \leq T_{{\rm IGM}}
$$
This is certainly a robust criterion, but it is possible that the actual 
suppression is much more efficient: if the IGM is at some finite temperature 
its contraction during gravitational collapse could be adiabatic and its 
temperature can rise. In a strict adiabatic collapse, one might 
expect that the actual suppression occurs on mass scales for which the 
temperature satisfies the following criterion:
$$
T_V \leq \Delta^{2/3}T_{{\rm IGM}}
$$

It is interesting that for $T_{{\rm IGM}} \sim 10^5 $ K, this leads to the 
suppression of structure formation for halos with $ V_c \sim 100$ km/s. 
However the existence of a cut-off for galaxy formation at a single 
circular velocity is not very attractive: observations do reveal the 
existence of small  galaxies and the luminosity function does not reveal 
any specific feature around $M \sim 16$. Actually, there might be an 
increase in the number of faint galaxies at this luminosity rather than 
a decrease as one would expect (ESO Slice Project, Zucca \etal 1998). 
Another  possible mechanism for suppressing galaxy formation is through 
photo-ionization. Not only does photo-ionization heat the gas to a 
temperature of the order of $T_{IGM} \sim 10^4 $ K, but it can also 
modify the cooling function. The implications are discussed in the next 
section.

\subsection{Effect of photo-ionization}

\begin{figure}  %[htbp]
\resizebox{\hsize}{!}{\includegraphics{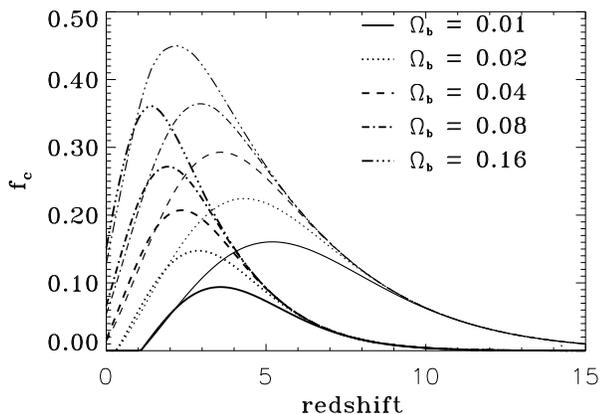}}
\caption{\label{fig:gzphot} Same quantity as in figure \ref{fig:gzomb} 
the instantaneous 
fraction of gas $f_c$  in two $\Gamma-$CDM pictures with $\Gamma = 0.5$ (thin 
lines) and 
$\Gamma =0.25$ (thick lines) with different values of the baryonic content of the universe 
with $b = 1.5$ in the photo-ionized IGM.}
\end{figure}

The existence of quasars at high redshift is enough to ensure that 
photo-ionization is playing a r\^ole in the history of the cosmological 
baryons. Whether the UV flux at high redshift  is high enough to ensure 
the low level of HI optical thickness observed towards high redshift quasars 
is still a matter of debate (Giallongo \etal 1996; Cooke \etal 1997; 
Devriendt \etal 1998).  In any case, photo-ionization 
by QSO's can easily heat the 
gas at a temperature of the order of $10^4$ K at redshift as high as 5. 
In addition photo-ionization suppresses 
collisional line cooling, by suppressing the existence of  neutral atoms 
(Efstathiou 1992). It is important to examine whether photoionization can severely limit the collapse of baryons within potentials as originally proposed by Efstathiou (1992). This has been examined in some details by mean of numerical simulations. Quinn et al. (1996) demonstrated that photoionization has essentially no effect on the collapse of halos for which the virial temperature is greater than $10^4$K. We can therefore compute  the amount of instantaneous cooling in a totally 
photo-ionized medium by  using  the cooling function of a fully 
ionized gas, assuming that only bremsstrahlung cooling is important 
(see for instance Thoul \& Weinberg 1996). 
The resulting $f_c$ is presented in figure \ref{fig:gzphot}. By comparing 
to figure \ref{fig:gzomb}, one can see that the effect of 
photo-ionization is to substantially suppress the amount of cooling occurring 
at low redshift.  The cumulative fraction of cooled gas
is then reduced by a factor of the order of 2 by redshift 0.
This implies that photo-ionization substantially changes 
the cooling process of the gas at low redshift, but can apparently not 
solve the overcooling itself. This is because the bulk of the overcooling   
occurs at a redshift when the density of the gas is high enough that 
photo-ionization cannot prevent most of the cooling.  Collin-Souffrin (1991)
has also examined the effect of photo-heating by the x-ray background, but 
this is not an efficient mechanism for raising the temperature much above 
few times $10^4$K. Our study indicates that photoionization does not solve the 
overcooling problem, i.e. that a large fraction of baryons collapse in small 
potentials at redshift below 10. This picture is rather different from what 
numerical simulations seems to indicate:  when photoionization is taken into 
account the emerging picture is that most of the IGM lies in moderately 
overdense region ($1+\Delta\sim 0.1-10$), and that no significant collapse 
occurs in small potentials. However, in order to see the collapse of baryons 
in the photoionized regime, great care must be taken to spatial and mass 
resolution 
(Weinberg et al, 1996). Let us examine this : present day  high resolution 
numerical simulations (see for instance Machacek et al., 1999) achieve 
typically $256^3$ cells with a box of the order 10$h^{-1}$ Mpc 
corresponding to a spatial scale of 40$h^{-1}$kpc (comoving). At redshift 
$z = 3$, this is the typical virial  radius  of a halo with a virial 
temperature of 
$T \sim 2. 10^5$K, significantly higher than the temperature  of photoionized 
gas. This artificial cut-off in the mass implies these numerical simulations 
cannot describe properly the formation of halos with virial temperature in the
 range $10^4-10^5$K, in which most of the cooling take place in the redshift 
range $3-10$. This suggests therefore that these numerical simulations  are 
still significantly limited by resolution, and for this reason do not see the 
overcooling problem. This is a serious problem for  the association of 
Ly$\alpha$ clouds to moderately overdense clouds might be an artifact due to 
insufficient resolution: the IGM might well be actually more clumpy than what 
is found in numerical simulations.

\begin{figure}  %[htbp]
\resizebox{\hsize}{!}{\includegraphics{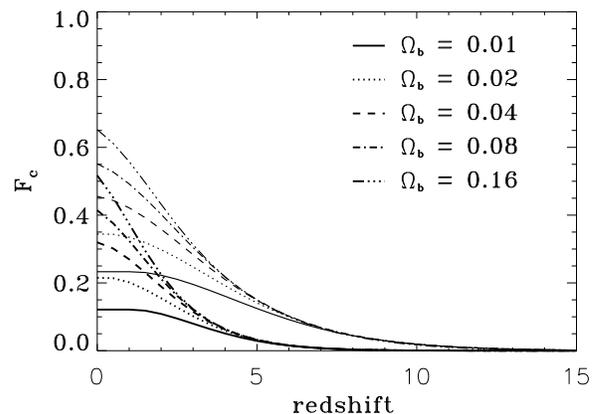}}
\caption{\label{fig:Fiphot} Same quantity as in figure \ref{fig:Fi1} 
the integrated 
fraction of gas $F_c$  in two $\Gamma-$CDM pictures with $\Gamma = 0.5$ 
(thin lines) and 
$\Gamma =0.25$ (thick lines) with different values of the baryonic content of the universe 
with $\eta = 1$ and $b = 1.5$ in the photo-ionized IGM.}
\end{figure}

\section{Galaxy formation in the warm IGM picture}

\subsection{Heating processes}

As photo-ionization seems not to be sufficient to solve the overcooling 
problem, one
has to advocate other heating processes.
Star formation is the source of a significant amount of energy liberation. Not 
only massive stars would radiate UV photons, but they would also end up in 
supernovae which could inject a significant amount of energy  in the IGM 
either through direct mechanical energy either via cosmic rays. The first 
possibility has been examined by Tegmark \etal (1993).  From theoretical 
considerations, it is possible that a significant  of energy produced by  a supernova is 
transferred to cosmic rays (Malkov \& Voelk 1995). This is also suggested 
by the observed amount of cosmic rays in the galaxy Drury \etal (1989). 
Prantzos \& Cass\'e (1994) 
have discussed possible evidence that the flux of cosmic rays was higher at 
the beginning of galaxy formation, in connection with the star formation 
process itself,  in order to explain the abundance of boron (see also
Nath \& Biermann 1993 and Ginzburg \& Ozernoi 1965). Both 
mechanisms can easily provide  heating sources for the IGM. Furthermore, 
cosmic rays may propagate rather easily to large distances, and therefore
 heat the IGM in a rather uniform way. It is therefore well 
possible that a substantial fraction of 
the energy produced by supernovae is transferred to the IGM. In the following 
we will therefore assume that a fraction of the energy produced by supernovae 
is transferred to the IGM. The energy from supernovae transferred to the IGM 
will therefore be written in a parametric way:
\begin{equation}
\dot{U}_{IGM}\|_+ = \epsilon \dot{\rho}_* {E}_{SN}
\end{equation}
where $ \dot{\rho}_* \sim 1/t_H(z)$ is the star formation rate per unit mass, 
${E}_{SN}$  is the energy 
produced by supernovae per unit mass of stars formed for a standard IMF
 and $\epsilon$ is 
the efficiency of the transfer mechanism. Assuming that the cooled gas 
is instantaneously transformed into stars , the heating source of the IGM then 
is:
\begin{equation}
\rho_b \dot{U}_{IGM} f_c
\end{equation}
Assuming a uniform temperature, $T_{IGM}$,   for the IGM containing all the 
baryons (i.e. neglecting the few baryons which are turned into stars), the 
equilibrium temperature in the self-regulated case  is given by the following 
equation:
\begin{equation}
(\rho_b/m_p)^2 \Lambda(T) = \rho_b \dot{U}_{IGM}  \frac{1}{\rho}\int_{m_1}^{m_2} N(m)m dm
\end{equation}
In this equation $m_1-m_2$ is the range of  masses in which cooling is able 
to occur. The lower temperature will be determined by the mass of the smallest objects able 
to form in a warm IGM. As we have argued, baryons will not collapse in 
potentials which 
end up  with a virial temperature smaller than the temperature of the IGM. 
On the other hand, as long as the IGM is in the cooling area,   
 the baryons will easily follow the collapse of the dark matter potentials 
provided that the cooling time of the gas is shorter than the dynamical time. 
 This means that the 
mass $m_1$ corresponds to the mass for which the virial temperature is equal 
to the temperature of the IGM. In the previous equation, $m_1$ therefore 
is a function of $T$, and the system is fully determined and hence
{\em self-regulated}: if overcooling occurs a large amount of stars will
 form, causing a large energy injection, thereby raising the temperature of
the IGM, suppressing galaxy formation by inhibiting further collapse. 
In practice, the system will evolve toward an equilibrium situation and the 
temperature can be computed. Such a solution may however not necessarily 
exist if the cooling is too efficient. This, for instance is the case at 
high redshift, when the number of heat sources is limited and the IGM is 
dense, so that cooling is very efficient.

\subsection{Supernovae Feedback}

We will try in the following to get some idea about the feed-back
parameter $\epsilon$, and about the subsequent production of metals
in the IGM.
It is known that the progenitors of the SNII are stars
of mass greater than $8M_{\odot}$ (Renzini \etal 1993). 
What we need to
compute is the mechanical (either thermal or kinetic) energy released
per unit mass of stars. This will provide  $ {E}_{SN}$,
the energy that the SNIIs might  release into the IGM.
We thus need to assume an Initial Mass Function (IMF) for star
formation, which we will take to be the Salpeter IMF:
\begin{equation}
\Psi (m) = A m^{-(1+x)}
\nonumber
\end{equation}
where $x=1.35$ and $\Psi(m)$ is the differential number of stars of
mass $m$.
We can then compute the ratio $p$ in mass of the SNIIs progenitors to
the total mass of stars:
\begin{equation}
p=\frac{\int_8^{100}m\Psi(m)\, dm}{\int_{0.1}^{100}m\Psi(m)\, dm} = 0.21
\end{equation}
for a Salpeter IMF. In the same way we compute the average mass of the
SNIIs progenitors $\langle M_{SNII}\rangle \simeq 30M_{\odot}$. 
Taking the average 
mechanical energy released by an SNII explosion to be 
$\langle E_{SNII}\rangle=10^{51}\,\rm{ergs}$, we can compute $\epsilon_0$:
\begin{equation}
{E}_{SN} = \frac{p\langle E_{SNII}\rangle}
{\langle M_{SNII}\rangle} = 3.5\,10^{15}\,\rm{ergs/g}
\end{equation}
Only a fraction $\epsilon$ of this energy will actually be available for 
heating the IGM. However, as we may consider a flatter IMF ($x=1.0$) or a 
bimodal one (Elbaz \etal 1995), it is possible in principle for 
this parameter to be greater than one.
For this reason, we will consider a range of models going from
$\epsilon=0.125$ to $\epsilon=2$.

\begin{figure}  %[htbp]
\resizebox{\hsize}{!}{\includegraphics{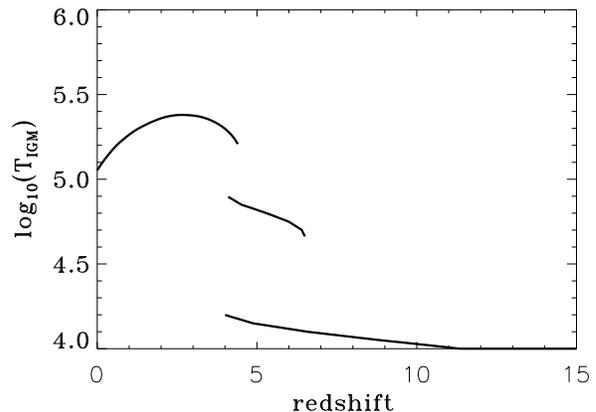}}
\caption{\label{fig:Tz1} An example of the equilibrium temperatures of the 
IGM in the self-regulated picture for two specific models : 
the parameter $\epsilon$ in two $\Gamma-$CDM 
pictures with $\Gamma = 0.5$ (thin lines) and 
$\Gamma =0.25$ (thick lines). The baryonic content of the universe 
is set to $0.1$, the normalization of the power spectrum is 
$b = 1.5$ and the feedback parameter is set to $\epsilon = 0.5$.}
\end{figure}

\begin{figure}  %[htbp]
\resizebox{\hsize}{!}{\includegraphics{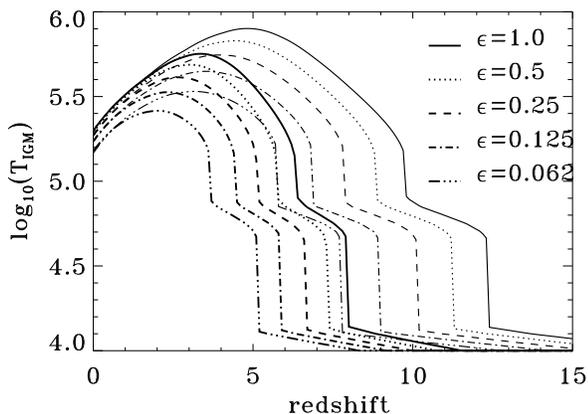}}
\caption{\label{fig:Tz2} The temperature of the IGM in the self-regulated 
picture for various values of the parameter $\epsilon$ in two $\Gamma-$CDM 
pictures with $\Gamma = 0.5$ (thin lines) and 
$\Gamma =0.25$ (thick lines). The baryonic content of the universe 
is set to $0.05$ and the normalization of the power spectrum is 
$b = 1.5$. }
\end{figure}

\subsection{ The warm IGM case}

In this section we will consider the case where only thermal energy
is injected into the IGM. As we already pointed out, in our scenario the only  free 
parameter, for a given IMF, is $\epsilon$, the efficiency with which energy from supernovae 
is transferred to the IGM.
The temperature of the IGM is then determined for each specific cosmological model. 
The general equilibrium solutions 
are   presented in figure \ref{fig:Tz1} for some models. 
Due to the various ``peaks'' in the cooling function several solutions
exist for the value of equilibrium temperature. Some of these solutions 
are unstable and are therefore rejected. One can see that no continuous 
solution exists over the whole range in redshift. In practice, the temperature
 has to ``jump'' from one 
equilibrium solution to another, a process known as a bifurcation. In 
such case, the exact 
behavior  of the temperature can be followed only by numerical integration of 
the non-equilibrium equations and is likely to depend sensitively on the 
details of the heating process, which is beyond the scope of this paper, since
here we focus on the general behavior. In the following, we will assume that
the temperature is always ``jumping'' to the highest temperature 
solution, which is the 
solution for minimal cooling, and is consequently the one 
with least heat input. The two branches exist over a limited range of redshift, so that this ambiguity has no consequences on the modeling.  
The temperature behavior is presented in figure \ref{fig:Tz2}. As shown there, 
the temperature rises above $10^4$K, between redshifts 5 and 15,
depending on the parameters of the model and reaches peak values of the order of a 
few $10^5$ K, that are in agreement with the Gunn-Peterson test. 
Interestingly enough, in the redshift range 2 to 5, the temperature of the 
IGM lies in a limited  range of values above $10^5$ K, rather independently of the details of the models. 
The existence of a warm IGM  is therefore a generic prediction of our scenario. 

%%%%%%%%%%%%%%%%%%%%%%%%%%%%%%%%%%%%%%%%%%%%%%%%%%%%%%%%%%%%%%%%%%%%%%%%%%%%%%%%%%%%
The instantaneous fraction of cooling gas, $f_c$, is expected to be turned
 into stars after some time and therefore should more or less represent 
the star formation rate. 
 In those models the cooling is spread over a wide range 
of redshifts and starts
at rather high redshift. Due to  the large suppression of the SFR this scenario already   possesses an interesting property: it leads to an 
integrated amount of stars by the present day which is in reasonably good agreement with observations as soon as the feedback mechanism is efficient at a level of 10\% (i.e. that 10\%
of the bulk of energy produced by supernova is injected in the IGM).\\. In such a model the overcooling problem is 
therefore essentially solved. This is a considerable success, given the limited number of free parameters of the model. However,
the star formation rate is found to be 
more or less constant with redshift, 
something which is not  in good agreement 
with recent data 
from the CFRS and HDF, which rather indicate late star formation (Madau, 1996). 

\subsection{Effects of combined photo-ionization and feedback}

Although we have shown that photo-ionization is clearly not a sufficient 
mechanism by itself to significantly  limit galaxy formation (at least in order to 
solve the overcooling problem in a realistic scenario), nevertheless
photo-ionization may play a significant role in the history of cooling process
(Efstathiou 1992). Furthermore, it is clear that photo-ionization is effective 
from a pure observational point of view, since the existence of 
quasars already provides a significant source  of UV photons at high redshift,
 even if they might
 not be entirely responsible for the ionization. We have 
therefore examined the case of an IGM which would be photo-ionized, and heated 
by some other mechanism (related to star formation) at the same time. The effect of 
photo-ionization is treated as in section~$3.2$, in which we use a simple 
cooling curve, assuming the ionization to be efficient enough to have 
completely suppressed collisional line cooling. \\

\begin{figure}  %[htbp]
\resizebox{\hsize}{!}{\includegraphics{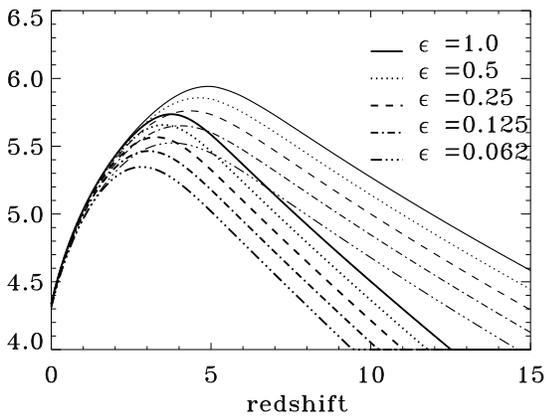}}
\caption{\label{fig:Tzphot} An example of the equilibrium temperatures of the 
IGM in the self-regulated picture for the case where the IGM is photo-ionized 
with an additional heating source for two specific models :
the parameter $\epsilon$ in two $\Gamma-$CDM 
pictures with $\Gamma = 0.5$ (thin lines) and 
$\Gamma =0.25$ (thick lines). The baryonic content of the universe 
is set to $0.05$ and the normalization of the power spectrum is 
$b = 1.5$. }
\end{figure}

The temperature has been computed as in the previous case. As photo-ionization renders 
the cooling curve regular, essentially by suppressing the line cooling due to 
bound atoms, bifurcations do not appear anymore. 
The resulting temperature history is given in figure~\ref{fig:Tzphot}. The 
behavior of the temperature is more regular than in the previous case, 
essentially due to the regular shape of the cooling curve. The temperature at 
high redshift is similar to the pure heating case (previous section), with value of the order of a 
few $10^5$ K, insensitive to the details of the 
model. However, there is  one noticeable difference at  low 
redshift : the temperature 
is decreasing faster, reaching values well below
$10^5$K. In our model, the temperature of the IGM at some epoch 
is directly indicative of the minimum mass of forming galaxies at this epoch.
The fact that the temperature reach lower values in the present case (see figure~\ref{fig:Tzphot},
compared to the pure heating case in figure~\ref{fig:Tz2}) is 
indicative that in a warm photoionized IGM {\em formation of smaller galaxies is allowed if
photo-ionization is efficient}  contrary to what one could naively expect.
\begin{figure}  %[htbp]
\resizebox{\hsize}{!}{\includegraphics{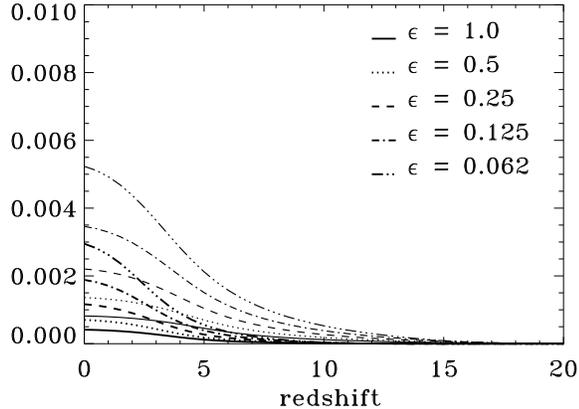}}
\caption{\label{fig:starTphot}   The integrated amount of stars in term of 
the critical density of the universe is given in two 
$\Gamma-$CDM pictures with $\Gamma = 0.5$ 
(thin lines) and 
$\Gamma =0.25$ (thick lines) with different values of the feedback parameter 
$\epsilon$ in the photo-ionized case, and $\Omega_b=0.05$.
}
\end{figure}
 
\section{Comparison with observations}

The essence of our model is that galaxy formation is suppressed at high 
redshift, because the IGM is pre-heated before collapse and that this 
warm IGM is maintained hot by self-regulation with ongoing star formation in 
early galaxies. The most direct observational
tests of this scenario would be provided by a measure of the temperature and the baryonic
content of the IGM at high redshift.
The best measurement of the Gunn-Peterson depth at high redshift 
(Giallongo \etal 1996), suggests that a photo-ionized IGM containing most of the 
baryons predicted by nucleosynthesis is ruled out. Numerical simulations
have however lead to a somewhat different picture in which most of the baryons 
lie in the Lyman-$\alpha$   systems. As we have discussed previously, these numerical simulations are probably limited by resolution. It is likely that dense regions form at high redshift which will still exist in the photoionized regime and in the reheated area and could then well lead to the Lyman-$\alpha$   systems.
This scenario has to be elaborated to see whether it can actually work, but  this is beyond the scope of this paper and will be the subject of a future paper. Limits on the Compton $y$-parameter 
can also provide an interesting upper limit on the pressure  of 
the IGM, but are still not stringent enough to constrain our scenario. A firm prediction of our model is that at redshift $\sim 2-4$ galaxies with circular velocity lower than 100 km/s do just not form, although they can form at higher and lower redshift. \\

\subsection{Neutral gas at high redshift}

The observations of quasars at high redshift have revealed the existence of 
damped Lyman-$\alpha$ systems, which are likely to be dense clouds of 
relatively cool gas. These clouds are believed to be the progenitors of 
present day disk galaxies. These clouds can be interpreted in our model as 
transient structures before they commence forming stars. Let   $t_g$ be 
the characteristic time of survival of the gas, if this time scale is  
short compared to the age of the universe $t_H(z)$ at epoch $z$,
then the amount of HI gas  can be related to $g(z)$ via:
\begin{equation}
\Omega_{\rm HI} \sim \frac{t_g}{t_H(z)} g(z) \Omega_{0}
\label{eq.Omg}
\end{equation}

\begin{figure}  %[htbp]
\resizebox{\hsize}{!}{\includegraphics{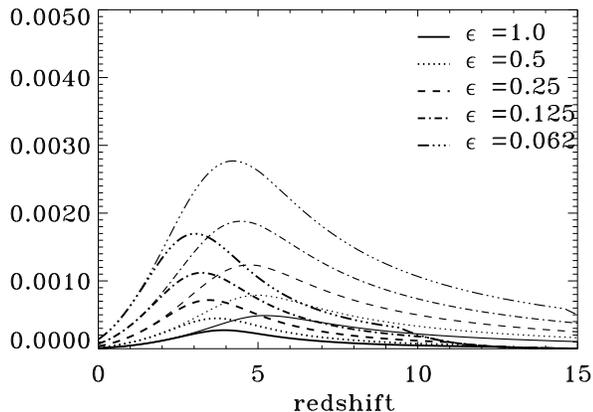}}
\caption{\label{fig.gHI}
The expected amount of HI against redshift in various models, with $\Omega_b=0.05$. Models and lines are the same  as in figure 
\ref{fig:Tzphot}.}
\end{figure}

In most scenarios,  star formation from HI gas is triggered by gravitational 
encounters, this would thus mean that the characteristic time scale $t_*$ should 
be of the order of the Hubble time. We have therefore estimated
$\Omega_{\rm HI} $ assuming the ratio $\frac{t_g}{t_H(z)}$ to be of the 
order of one. The results are presented in figure~\ref{fig.gHI}.
Our models exhibit the correct qualitative behavior for the 
evolution of the HI content.  For the  CDM picture
with $\Gamma = 0.5$, the maximum is located
at a redshift slightly higher than observed, but with $\Gamma = 0.25$ 
the shape is in satisfactory agreement with the data.  \\

In order to build a consistent picture we select as an optimum case
only models which reproduce both the correct density of stars, and 
the observed density of HI as a function of redshift (Lanzetta \etal 1995;
Storrie-Lombardi \etal 1996; Natarajan \& Pettini 1997). Models which 
lead to a reasonable $\Omega_*$ are not far from reproducing correctly 
the amount of observed HI. This is not surprising, given the fact that 
the observed HI is known to be more or less of the order of what is required to
explain present day stars. It is also important to notice that this therefore justifies
 our assumption $\frac{t_g}{t_H(z)} \sim 1$ : using a smaller 
value would lead to an overproduction of stars. The photo-ionized
 case with feedback succeeds particularly well to match the data for 
$\Gamma = 0.25$, while the case with $\Gamma = 0.5$ fails to fit. 
This is of interest since this offers a constraint on the power spectrum 
on scales much smaller than usually constrained 
(from clusters or CMB data for instance), still leading to a preferred 
value of  $\Gamma = 0.25$. The pure heating case also reproduces 
the observations reasonably, but not in detail: the amount 
of HI gas at low redshift is over-produced by a  factor of two, while 
a substantial amount of HI gas appears at high redshift ($z > 4$)
although this might not be regarded as a real problem given the absence of 
information at such redshift.\\

 The fact that our model reproduces so naturally the observed amount of HI 
 is probably its most successful feature: in fact, we found that a $\Gamma = 0.25$ 
photo-ionized warm IGM model reproduces 
 automatically the observations as soon as 
the feedback parameter is tuned to reproduce the present-day 
observed amount of stars, a non-trivial result.

\begin{figure}  %[htbp]
\resizebox{\hsize}{!}{\includegraphics{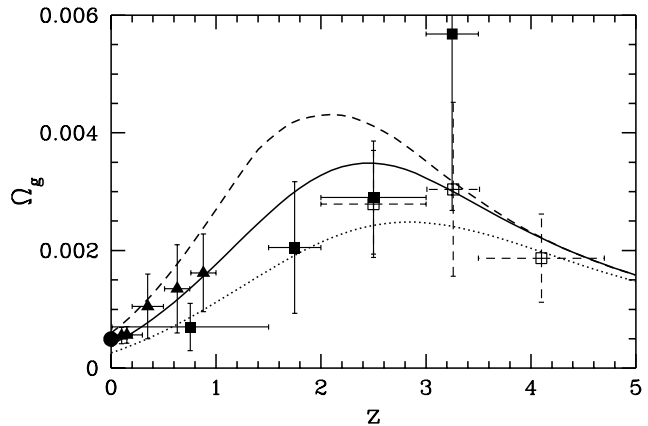}}
\caption{\label{fig.gHIphot}
The amount of HI against redshift in  models with parameters selected to
reproduce  quite well the observed $\Omega_*$ with $\Gamma = 0.25$ in the warm 
photo-ionized case. {\em Dotted}, {\em solid} and {\em dashed} lines correspond
respectively to the models with ($\Omega_b$,$\epsilon$) equal to ($0.05$,$0.125$),
($0.1$,$0.5$) and ($0.2$,$2.0$).
The $\Gamma = 0.5$ case never fit properly the data.
Data points are taken from Rao \& Briggs 1993 ({\em filled circle}), 
Natarajan \& Pettini 1997 ({\em filled triangles}), Lanzetta \etal 1995
({\em filled squares}), and Storrie-Lombardi \etal 1996 ({\em open squares}).}
\end{figure}

\begin{figure}  %[htbp]
\resizebox{\hsize}{!}{\includegraphics{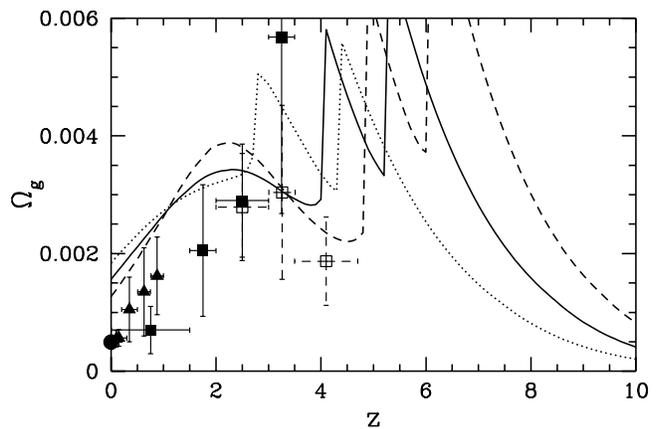}}
\caption{\label{fig.gHInophot}
Same as figure~\ref{fig.gHIphot}, but without photo-ionization.
The $\Gamma = 0.5$ case never fits properly the data.}
\end{figure}

\subsection{ Cosmological star formation history}

Given the success of the model thus far, it is tempting to go one step 
further and see whether the model might reproduce other features of the 
cosmic baryon history. One of the most important advances in recent years 
has been the estimation of the cosmic star formation history based on the 
CFRS and HDF data. Although star formation is certainly a very complex process
it is possible that simple  assumptions may lead to a reasonable picture. 
Semi-analytical models have attempted a first step in this direction, 
but  based
on a rather different approach than ours. A 
realistic picture would need a more detailed inventory, with attention 
paid to various types of galaxies, merging trees, etc.  
%However, here we want 
%to catch the basic behavior of the model, and will not go in such a high level
%of elaboration, that we will leave for future works.   
In order to investigate the star formation history in our model, we have therefore 
 adopted a simple rule : the fraction of gas available at one epoch is assumed 
to lead 
to star formation during a fixed period of time $\Delta t_*$. This time scale 
represents the typical time scale for consumption of the gas that 
will be turned into stars, which could be somewhat different from the survival 
time scale of the HI gas ($t_g$) that we introduced previously.  The cosmic 
star formation rate at an epoch is therefore:
\begin{equation}
\dot{\rho} \sim \frac{1}{t_H(z)} \int_{t_H(z)-\Delta t_*}^{t_H(z)} 
\Omega_{b}g(z)dt
\label{eq:csr}
\end{equation}

\begin{figure}  %[htbp]
\resizebox{\hsize}{!}{\includegraphics{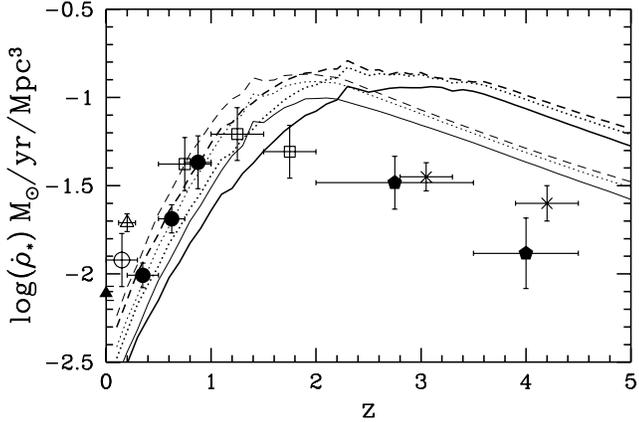}}
\caption{\label{fig.retard}
Star formation rate estimated by Eq. \ref{eq:csr} for two value of the 
time delay for gas consumption into stars. The thin lines are for 
$\Delta t_*= 1$Gyr, and the thick lines are for $\Delta t_*= 2$Gyr. 
The data points are taken from Lilly \etal 1996 ({\em filled circles}),
Madau \etal 1996 and 1997 ({\em filled pentagons}), Steidel \etal 1998 ({\em crosses}), 
Conolly \etal ({\em open squares}), Gallego \etal ({\em filled triangle}),
Tresse \& Maddox 1998 ({\em open triangle}) and Treyer \etal 1998 ({\em open circle}). 
The parameters used are those already used in figure \ref{fig.gHIphot}.}
\end{figure}

We have considered two values of $\Delta t_*$, 1 and 2 Gyr. The resulting 
cosmic star formation history is plotted in figure~\ref{fig.retard}. Good 
agreement is found in the low redshift regime. This agreement is closely 
connected to the inclusion of photo-ionization. The pure heating case 
does not match the data well.  Although the modeling of cosmic star formation
 history is likely to be much more complex than the simple ansatz we have followed, it is nevertheless interesting to see how well 
the prediction matches the observations.

\subsection{Metal enrichment}

The subject of metal enrichment of the Intra-cluster Medium (ICM)
 has been investigated by
numerous authors (e.g. see
Renzini \etal 1993; Renzini 1997; Mushotzky \etal 1996; 
Ishimaru \& Arimoto 1997; Gibson \etal 1997; Fukazawa \etal 
1998); and it has been
advocated that the ICM could be representative of the IGM as far as
metals are concerned (Renzini 1998). Self-consistent treatments
 of the metallicity of the IGM in spectrophotometric models of galaxies
are just in their infancy (Sadat \etal 1999). As  the quantity of iron
produced by SNII is proportional to the energy released in the
IGM, we can estimate the abundance of iron from SNII in the IGM in our model:
\begin{equation}
Z_{SNII}(z=0) =0.5\times B_*(z=0) p\langle M_{Fe}\rangle 
/\langle M_{SNII}\rangle
\end{equation}
where $B_*(z=0)$ is the present fraction of baryons turned into stars,
$\langle M_{Fe}\rangle$ is the average iron mass produced by a
typical SNII explosion, and the factor $0.5$ accounts for an
equipartition of the iron between the stars and the IGM. Adding the contribution of the SNIas which  provide 75\% of the iron in the IGM (see Renzini et al., 1993):we get:
\begin{equation}
Z_{tot}(z=0)=Z_{SNII}+Z_{SNIa}=4\times Z_{SNIIs}(z=0)
\end{equation}
One thus finds typical metallicities of the IGM of $0.1$~solar 
at redshift $z=0$,
and of $0.05$~solar at redshift $z=2.5$, which are compatible with
recent observations in Damped Lyman~$\alpha$ Systems (Lu \etal 1996; 1997),
these systems are assumed to be representative of the state of the IGM as
far as metals are concerned. This is not a strong constraint on our model however, as the metalicity of the IGM essentially reflects the metal production 
inferred from the star formation rate.

\subsection{Discussion and Conclusion}

We have examined in detail the overcooling problem once again.
We confirm the previous claim that in the absence of significant feed-back mechanism,
cold gas, eventually in the form of stars, is over-produced in Cold Dark Matter models.
Observations reveal a little amount of stars compared to standard nucleosynthesis
prediction. A fundamental question is therefore: where are the primordial baryons ?
There are reasonable arguments which indicate that they are in the intergalactic 
gas, maintained at some temperature above $10^4$K. We found that photo-ionization 
reduces significantly the amount of cooling, especially in the case of a low baryonic density.
However, from our analysis,
we found that the photo-heating 
is not sufficient to suppress the cooling of the gas, over-producing in a significant way 
the observed
amount of cold gas. This strongly argues for the existence of strong feed-back
 mechanisms
during galaxy formation. We have explored in some
detail the warm IGM picture in which the high redshift gas is heated preventing 
further galaxy formation. We found that the IGM can be easily heated to temperatures 
of the order of $10^5-10^6$K, altering
substantially the process of galaxy formation. In such models, the overcooling problem
is easily solved provided that at least 10\% of the energy of supernova is transferred
to the IGM. We found that with this hypothesis the observed amount of 
stars can be reproduced for very reasonable values of the feedback parameter.
Such scenarios reproduce quite well the observed amount of HI gas at high redshift,
especially when photo-ionization is taken into account.  This is a remarkable success of this
model. Assuming that this gas is turned into stars, we also found that the cosmic stars 
formation rate is well reproduced. Clearly, this model seems to meet success, despite the fact 
that the galaxy formation history is significantly different than in more traditional scenario
 in which large galaxies are built from the merging of smaller entities. This certainly 
illustrates the fact that large galaxies may well have been formed at relatively high redshift
(say between 3 and 5), in $\Gamma$--CDM model with $\Gamma= 0.25$. High redshift galaxy formation 
of large galaxies is not therefore intimately linked to low density universes (although very high 
redshift - greater than 10 - would certainly be a problem in a high $\Omega $ universe).
 
It is interesting to notice at that point that this scenario provides us also with a natural explanation
for the $L_X-T_X$ departure from scaling law for small clusters of galaxies (Kaiser 1991; 
Evrard \& Henry 1991). Indeed, if the IGM is pre-heated before the formation 
of the clusters, then the zero-point
entropy of the gas prevents it to be very concentrated in the cores of clusters. This effect is
very important when this zero-point entropy is comparable or bigger than the entropy created
by shocks during the formation of the cluster, and then affects primarily small galaxies, clusters and
groups. This departure from pure scaling can also be seen in the evolution of the surface brightness
profiles for different gas temperature ranges (data taken from ROSAT and GINGA, Ponman \etal 1998).
These authors estimate this zero-point entropy to be of the order of $100\,h^{-1/3}\,{\rm keV\,cm}^2$,
which can be provided by a uniform IGM pre-heated to a temperature $T_{IGM} \sim 3\,10^4(1+z)^2\,
{\rm K}$. The required temperature is then $\sim 2.5 \, 10^5\,{\rm K}$ at $z=2$, which is compatible 
with the results of our model (see Fig.~\ref{fig:Tzphot}).\\

In our scenario, galaxy formation occurs in two clearly distinct phases.
At high redshift objects which form are cooling with a cooling time much shorter
than the age of the universe. This could be argued as being the phase of bulges and 
elliptical formation. In this regime, the formation history follows the classical hierarchical 
picture, small objects form first, large objects latter. At later epochs, this scheme is reverted.
This might well be the epoch of disk formation, during which the gas gently falls in some potentials
which already contain stars from  the previous generation. This would imply that the
typical epoch for disk formation is $z = 0$ for the smallest to $z = 2-3$ for the largest disks.
Beyond this epoch most of the forming structures would correspond to bulges and ellipticals.
This scheme is certainly consistent with observations as high redshift objects seem to 
have characteristics of bulges in their early stages. A firm prediction of our model is that galaxy formation with circular velocity smaller than 100 km/s should be strongly suppressed at redshift in the range $1-5$. Clearly, observations of the dynamical state
of high redshift galaxies would be important to test this scenario. This is probably testable in future NGST observations. Of course, the most direct test 
of our scenario would be to measure the temperature history of the IGM. However, this seems to be 
difficult.

\begin{acknowledgements}
SP would like to thank Priyamvada Natarajan for a careful reading of the manuscript.
This work is partially supported by NSERC of Canada.
\end{acknowledgements}

\end{document}